\documentclass[amssymb,prb,twocolumn,floats,amsmath,showpacs]{revtex4}
\usepackage[T1]{fontenc}
\usepackage{bm}
\usepackage{graphicx}% Include figure files
\usepackage{amssymb}
\usepackage{leftidx}
\usepackage{upgreek}
\usepackage{siunitx}
\usepackage{amsfonts}
\usepackage{amsmath} 
\usepackage[outdir=./]{epstopdf}
\usepackage{color}

\begin{document}

\title{Hybrid electroluminescent devices composed of (In,Ga)N micro-LEDs and monolayers of transition metal dichalcogenides}
\author{K.~Oreszczuk$^1$, J. Slawinska$^2$, A. Rodek$^1$, M. Potemski$^{1,3}$, C. Skierbiszewski$^2$, and P.~Kossacki$^1$} 

\address{$^1$  Institute of Experimental Physics, Faculty of Physics,
University of Warsaw, Pasteura 5, 02-093 Warsaw, Poland}
%\address{$^2$ Institute for Quantum Electronics, ETH Z\"{u}rich, CH-8093 Z\"{u}rich, Switzerland}
\address{$^2$ Institute of High Pressure Physics, Polish Academy of Sciences, Soko\l owska 29/37, 01-142 Warsaw, Poland}
\address{$^3$ Laboratoire National des Champs Magn\'etiques Intenses,
CNRS-UGA-UPS-INSA-EMFL, 25 rue des Martyrs, 38042 Grenoble, France}

\begin{abstract}
We demonstrate a novel electro-luminescence device in which GaN-based $\upmu$-LEDs are used to trigger the emission spectra of monolayers of transition metal dichalcogenides, which are deposited directly on $\upmu $-LED surface. A special $\upmu$-LED design enables the operation of our structures even in the limit of low temperatures. A device equipped with a selected WSe$_2$ monolayer flake is shown to act as a stand-alone, electrically-driven single-photon source.
\end{abstract}
\maketitle

\maketitle

\section{Introduction}
Atomically thin semiconductors, such as monolayers of transition metal dichalcogenides (TMDs) and their homo- and hetero-structures, exhibit interesting optical properties. The list of prominent attributes of these two-dimensional (2D) structures includes: strong excitonic binding\cite{you2015, he2014, ross2013} which favors room temperature capabilities\cite{lorchat2018ACSPhotonics}, possibilities to manipulate the valley degree freedom \cite{mak2010, splendiani2010, zeng2012, mak2012control}, and emergence of new class of quantum emitters\cite{koperski2015, tonndorf2015, brooks2018theory, kern2016goldrods, kumar2015dwuwarstwa, branny2017, luo2019160k}. In parallel to the intense fundamental research, pertinent efforts are also undertaken to exploit the TMD semiconductors in designing and fabricating of new optoelectronic devices of possible practical use\cite{blauth2018NanoLett.a, peyskens2019NatCommun}. The operation of such devices is desired to be electrically driven. Controlled doping of TMD materials, although essential for the operation of conventional electroluminescent p-n diodes, still remains a largely unresolved challenge \cite{wang2018Adv.Mater., qiu2013NatCommun, wang2016Phys.Chem.Chem.Phys.}.

 Therefore, alternative methods are being developed, relying most often on tunneling mechanisms of carrier injection into the active 2D component of the device. In fact, the electrically driven emission from TMD semiconductors has been successfully demonstrated in a number of differently designed structures \cite{palacios-berraquero2016NatureCommunicationsa, sundaram2013NanoLett., ye2014Appl.Phys.Lett., cheng2014NanoLett., withers2015NatureMater, wang2017NanoLett., li2015NatCommun}. However, the architecture of such structures implies rather complex manufacturing processes, and the performance of the reported devices remains not yet fully satisfying. This calls for further efforts to improve/optimize the already proposed device schemes and/or to search for optional solutions.

Taking advantage of the advanced technology of nitrides\cite{wu2018Appl.Sci.}, we put here forward a new concept of a compact pseudo-electroluminescent device by integrating the TMD semiconductors into (In,Ga)N light-emitting diodes (LEDs). Our devices are composed of micro-(In,Ga)N-LEDs ($\upmu$-LEDs), on top of which we directly deposit the TMD layers: the electrically driven (In,Ga)N $\upmu$-LEDs serve as the excitation sources to generate the PL emission from TMDs. A special design of our (In,Ga)N diodes ensures their functionality at cryogenic temperatures, which are preferential conditions for generating light from TMD monolayers and an unavoidable requirement when tracing the quantum emission centres from these layers. Low-temperature operation of our electroluminescent devices comprised of different: MoS$_2$, MoSe$_2$, WSe$_2$ and WS$_2$ monolayers, is demonstrated. Importantly, the hybrid (In,Ga)N-$\upmu$-LED/WSe$_2$-monolayer device is also shown to function as an electrically driven source of single photons. While the fabrication of the present devices implies a laborious exfoliation manufacturing, we speculate that the industrial scalability of such devices could be approached in the future, taking into account the progress in the growth of TMDs by epitaxy methods and thus evoking a possibility to grow the TMDs directly on (In,Ga)N LEDs, particularly with molecular beam epitaxy techniques\cite{pacuski2020NanoLett.a}.

The technology of nitride LEDs has undergone intense development in recent years\cite{wu2018Appl.Sci.}. Relevant efforts have been made to respond to pertinent demands for low energy consumption, that is, to achieve high luminous efficacy of LEDs, but also to improve their brightness and contrast, their shock resistance, and degradation time \cite{wu2018Appl.Sci.}. Although many nitride LEDs already find relevant applications in display technology \cite{day2011Appl.Phys.Lett., chen2021J.Phys.DAppl.Phys.}, the new possibilities for nitride optoelectronics have been raised more recently by employing a scheme of a tunnel junction (TJ) \cite{skierbiszewski2018201876thDeviceRes.Conf.DRC, kowsz2017Opt.ExpressOE, skierbiszewski2018Appl.Phys.Express}. Such architecture of (In,Ga)N LEDs is also crucial for the present work. The buried TJ enables the nitride LEDs to operate at cryogenic temperatures\cite{chlipala2020Opt.ExpressOE}. Equally important is that the use of the TJ allows us to replace the top p-type layer (of a conventional LED) with a highly conductive n-type layer. This enhances the current spreading and enables the n-type application of the side contacts to the device. Thus, TMD flakes can be directly deposited on the GaN surface, and the experiments can be carried out at low temperatures.

\begin{figure*} \begin {center} 
\includegraphics[width=180mm]{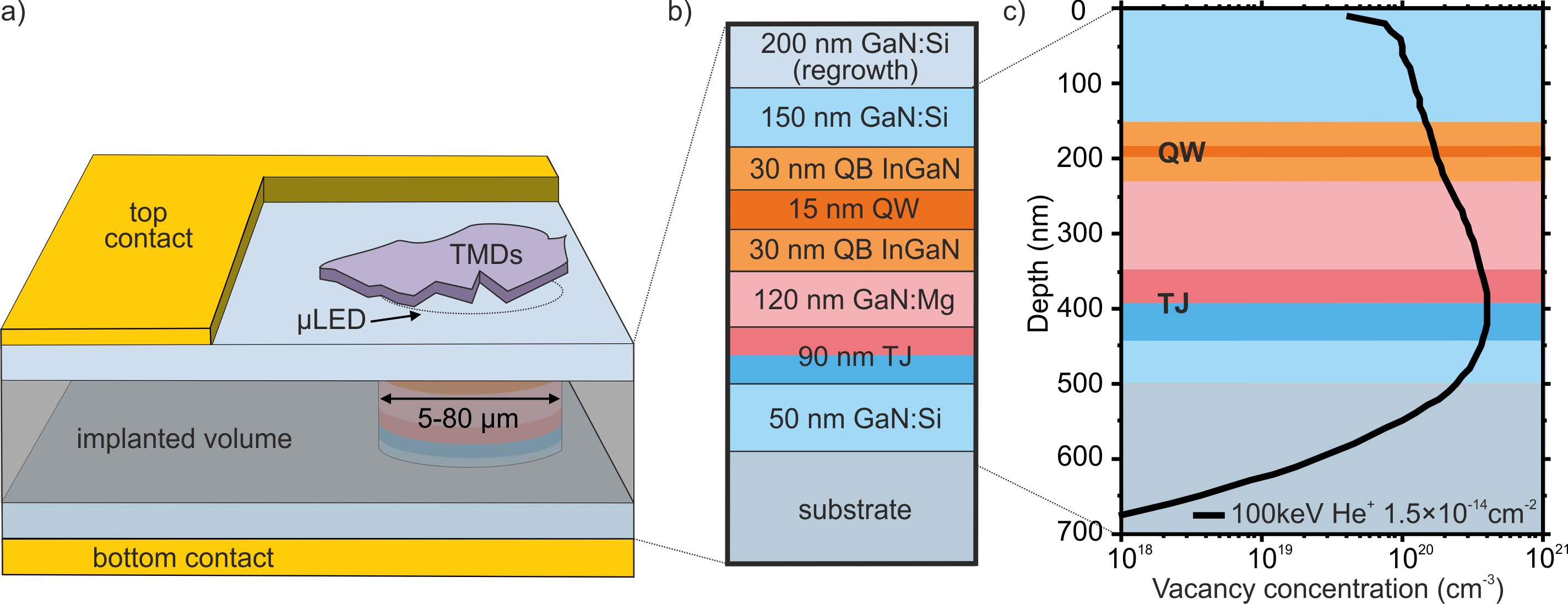}
\caption{ (a) Schematic of the hybrid (In,Ga)N $\upmu$-LED/TMD-monolayer device. (b) Sketch of the nitride $\upmu$-LED structure grown by plasma assisted molecular beam epitaxy, with a characteristic sequence of epitaxial sections. (c) Ion-implantation profile for the implemented He$^+$ energy of $100$\,keV and dose of $1.5\cdot10^{14}\,$cm$^{-2}$.}
\label {led} \end {center}\end {figure*}

\section{Experimental details}

The epitaxial structure, schematically shown in Figure \ref{led}, was grown by plasma-assisted molecular beam epitaxy (PAMBE) on (0001) bulk GaN crystals. The PAMBE grown TJ ($90$\,nm) and $\upmu$-LED structure that emits light at a wavelength of $450$\,nm was capped with the n-type GaN of $150$\,nm (Figure \ref{led}b). The emission area of $\upmu$-LEDs was defined by a photolithography mask and He$^+$ implantation ($100$\,keV, $D=1.5\cdot10^{14}\,$cm$^{-2}$) of the $\upmu$-LED structure. Ion energy and dose were adjusted to penetrate the region at the location of the TJ (Figure \ref{led}c). The vacancies produced during the ion implantation of the TJ deteriorate the p-type and n-type conductivity and significantly increase the resistance of the TJ in the areas outside the $\upmu$-LEDs. After the ion implantation, the regrowth of $200$\,nm of high conductive n-type GaN enhances the current spreading at the top of $\upmu$-LEDs and allows the creation of the side contacts to the device. After regrowth, the $150\times 300\,\mu$m devices containing arrays of individual microdevices were separated by reactive-ion etching. Next, on the Ga-polar side of the samples, the Ti/Al/Ni/Au metal contact was deposited with a photolithographic mask, followed by lift-off. The same metal contacts were used for bottom contact on the nitrogen polar side.

The TMD monolayers have been placed on top of (In,Ga)N $\upmu$-LEDs (see Figure \ref{led}a) using deterministic transfer procedures. Monolayer flakes were mechanically exfoliated from bulk crystals with chemically pure backgrinding tape and transferred to the $\upmu$-LED surface using Gel-Pak DGL-X4 elastomeric films. A number of devices have been fabricated, comprising different, MoS$_2$, MoSe$_2$, WSe$_2$ and WS$_2$ monolayers.

The optical measurements were carried out in the two experimental configurations. Time-integrated studies were performed with the samples placed in the cold-finger cryostat. The emission of TMD monolayers has been measured in two different configurations. Primarily, the measurements have been carried out using the external laser excitation at 532 nm  ($2330$\,meV). Secondly, the luminescence was triggered from underneath with (In,Ga)N $\upmu$-LEDs. 

The spectra composed of sharp emission lines characteristic of a WSe2 monolayer have also been examined with photon correlation experiments. These experiments have been carried out in the Hanbury-Brown-Twiss configuration, under (In,Ga)N LED excitation, and required the device to be immersed in superfluid helium to minimize the undesirable effects of heating (when large currents are driven to trigger the (In,Ga)N LEDs).

\begin {figure} \begin {center} 
\includegraphics{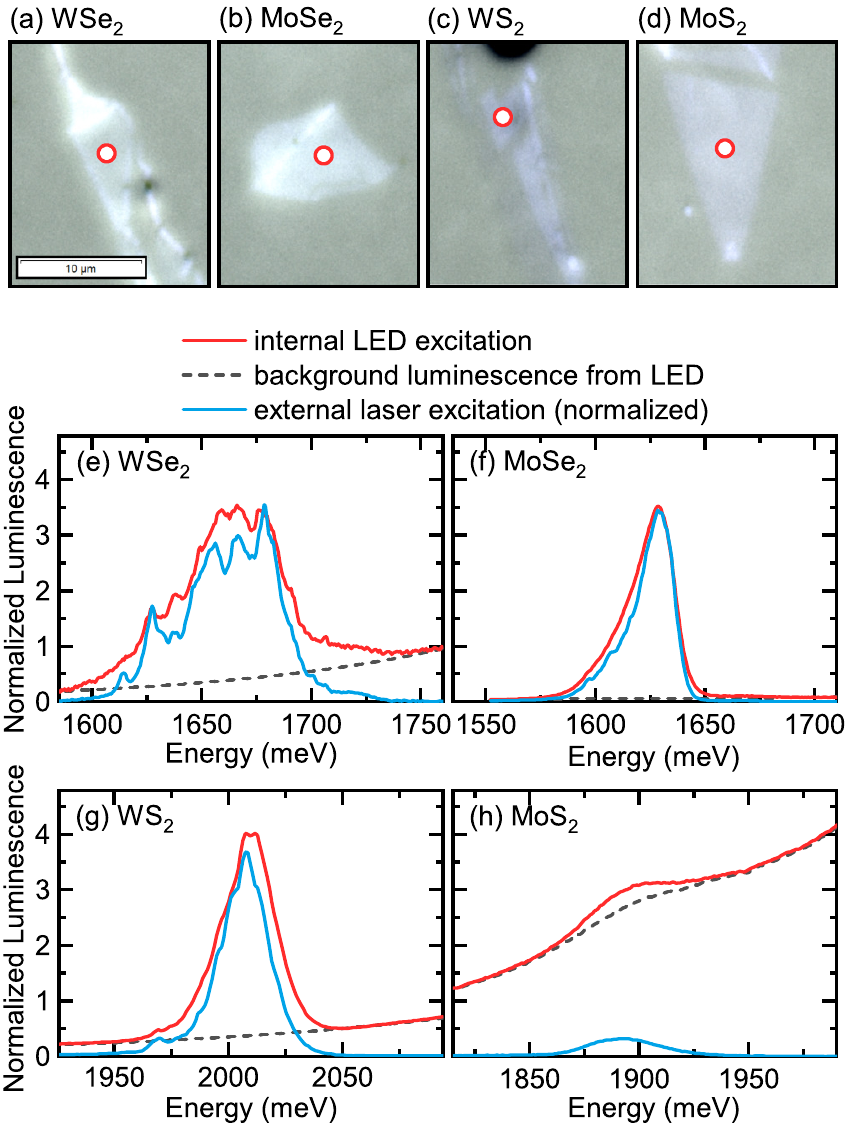}
\caption{ (a-d) Images of the four different monolayers of TMDs. The flakes are exfoliated on $\upmu$-LEDs, which span the whole photo area. The $\upmu$-LED current density is equal to $0.3\,\mu$A$\,\mu$m$^{-2}$. Red circles mark the area from which the luminescence was acquired. (e-h) Spectra of the monolayers acquired under $\upmu$-LED excitation (red) and $532\,$nm ($2330\,$meV) external laser excitation (blue). Dashed line: $\upmu$-LED luminescence acquired at the off-flake spot on the sample.}
\label {monolayers} \end {center}\end {figure}

\section{Results and discussion}
 The optical microscopy images of four different, WSe$_2$, MoSe$_2$, WS$_2$ and MoS$_2$ monolayers deposited on top of individual (In,Ga)N $\upmu$-LEDs are illustrated in Figure \ref{monolayers}a-d. The characteristic, low temperature ($10\,$K) microluminescence spectra of these layers are shown, correspondingly, in Figure \ref{monolayers}e-h. The spectra shown with blue traces were measured under excitation with the external laser ($532$\,nm), whereas those shown in red have been activated by the underneath (In,Ga)N $\upmu$--LEDs. The photoluminescence signals measured under external excitation are typical (with respect to spectral shape and characteristic energies and intensities) of TMD monolayers, as often studied when they are directly deposited on conventional Si/SiO$_2$ substrates. Characteristic spectra of our WSe$_2$, MoSe$_2$ and WS$_2$ monolayers are fairly well reproduced when they are excited with (In,Ga)N $\upmu$-LEDs. On the other hand, the apparent background signal (see dashed traces in Figure \ref{monolayers}e-f), which is due to an unfiltered low-energy tail of the (In,Ga)N emission, largely distorts the spectrum of the MoS$_2$ monolayer. This is, in fact, because of a rather weak luminescence efficiency of MoS$_2$ monolayers, as commonly reported when this particular TMD monolayer is not protected by hexagonal boron nitride.

 \begin {figure} \begin {center} 
\includegraphics[width=86mm]{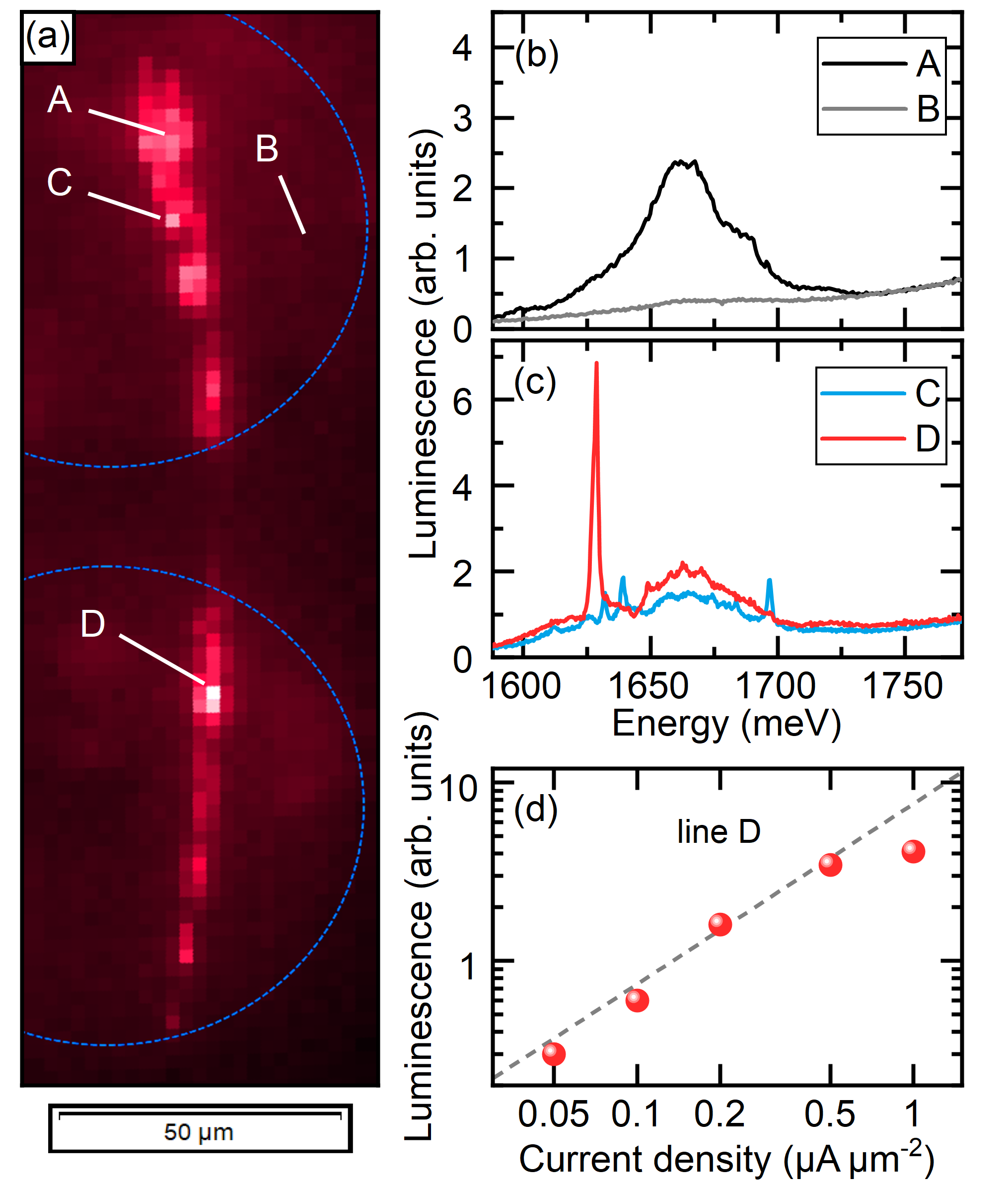}
\caption{ (a) Map of the luminescence of a long and narrow flake of a monolayer WSe$_2$. The signal is integrated over $1625$\,meV - $1640$\,meV range, at the $\upmu$-LED current density of $0.5\,\mu$A$\,\mu$m$^{-2}$. Blue circles mark the outlines of the two circular $\upmu$-LEDs. Symbols A-D mark the spots presented in panels (b-c). (b) $\upmu$-LED excited PL of the homogeneous part of the WSe$_2$ flake (black) and the background luminescence of the $\upmu$-LED (grey). (c) $\upmu$-LED excited PL from the two spots with the narrow-line emitters. (d) The dependence of the luminescence intensity of line D on the $\upmu$-LED current density. The excitation sublinearity is caused by the elevated sample surface temperature at the highest current densities. }
\label {map} \end {center}\end {figure}

Having established the overall functionality of our pseudo-electroluminescence devices, we now turn our attention to structures displaying sharp emission lines, possibly comprising the quantum emitter centers. Such centers preferentially appear in WSe$_2$ layers. Several $\upmu$-(In,Ga)N-LED/WSe$_2$-monolayer devices have been fabricated and tested. The irregular shape of WSe$_2$ flakes has been found to favour the observation of sharp emission lines. One may speculate this is due to the increased probability of the formation of the specific centers trapping the photoexcited carries at the edges of TMD flakes\cite{koperski2015, tonndorf2015} or induced by inhomogenous strain distribution\cite{brooks2018theory, kern2016goldrods, kumar2015dwuwarstwa, branny2017, luo2019160k}.

The representative example of a device displaying sharp emission lines contains a narrow, elongated flake of a WSe$_2$ monolayer. The device has been mounted on a cold finger of the cryostat, cooled down to $10$\,K and primarily tested under such conditions. After turning on the (In,Ga)N diodes, the emission spectra, in the spectral range $1590-1775$\,meV, were measured when scanning the surface of the WSe$_2$ flake and its surrounding, with a spatial resolution of $2\,\mu$m. A map of the intensity of the PL integrated over the spectral range of  $1625-1640$\,meV is shown in Figure \ref{map}a. This image illustrates well the shape of the flake, which extends over the active areas of two neighbouring (In,Ga)N $\upmu$-LEDs. The background signal due to the emission tail of the (In,Ga)N diode is fairly weak - see the spectrum measured outside the flake area (Figure \ref{map}b). The spectra measured in the areas in the middle of the flake, away from its edges, exhibit a broad feature, centered at $1660$\,meV --  a typical response of WSe$_2$ ML with significant spatial inhomogeneity (see Figure 3b). However, near the edges and narrow parts of the flake, the luminescence spectra often display sharp lines; see Figure \ref{map}c. 

Notably, the low activation current (below $1\,$nA$\,\mu$m$^{-2}$) of our device permits the measurements of the emission spectra of the WSe$_2$ monolayer over a broad range of current densities driving the (In,Ga)N diodes and thus enables the straightforward tuning of the excitation power to match the saturation point of a particular emitter. To illustrate this effect, we present a typical PL intensity of the emitter as a function of the current density (Figure \ref{map}d). As the luminescence intensity of narrow lines in WSe$_2$ monolayers is known to drop down upon increase of temperature\cite{oreszczukprb} we conclude that the observed effect is likely due to a low cooling efficiency of our device under the current arrangement (limited by the heat transfer effectiveness between our device and the mounting base in the cryostat).
However, the heat generation process can be suppressed by reducing the size of the $\upmu$-LED and the precise positioning of the TMD flakes.

To demonstrate the operation of our device as a single-photon source, the device was immersed in a superfluid helium bath and cooled down to a temperature of $1.65$\,K. This resulted in an improved cooling efficiency of the device, enabling the application of higher diode currents, and at the same time gaining higher emission intensities of narrow emission lines. This allowed for the correlation measurements of the photon statistics emitted from the particular centre. 

\begin {figure} \begin {center} 
\includegraphics{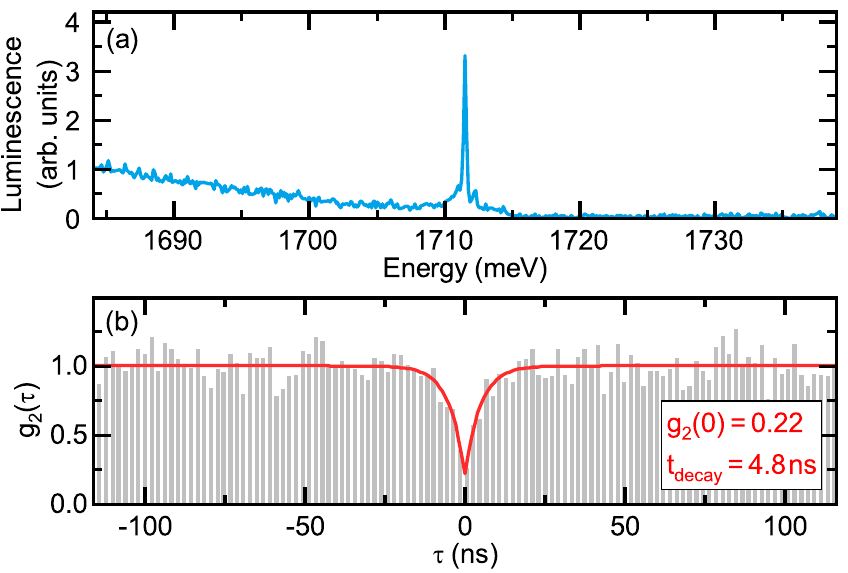}
\caption{ (a) Luminescence spectrum of the narrow emission line chosen for the single-photon correlation experiment. (b) Coincidence correlation counts as a function of the time interval between photons. The photoluminescence is excited with $\upmu$-LED. The broad background luminescence from the $\upmu$-LED was not subtracted in the post-processing.}
\label {correlation} \end {center}\end {figure}

In Figure \ref{correlation}a we present the spectrum of the narrow line emitter in the WSe$_2$ monolayer under the (In,Ga)N $\upmu$-LED excitation that was powered with $1.8\,\mu$A$\,\mu$m$^{-2}$ current. We note the particularly low ratio of the background $\upmu$-LED spectral tail to the narrow emission line intensity. Any uncorrelated photons detected in the Hanbury-Brown-Twiss experiment would proportionally reduce the amplitude of coincidence dip as well as the feasibility of our device as a single-photon source.

The acquired photon coincidence correlation function, $g_2(\tau)$, is presented in Figure \ref{correlation}b. The photon coincidence correlation function $g_2(\tau)$ denotes the relative probability density of an event with a time separation $\tau$. The decrease in the $g_2$ value around $\tau = 0$ evidences the single-photon component in the emission. The emission has predominantly single-photon characteristics  when $g_2(0) < 0.5$. 

The amplitude of the correlation function at zero delay is significantly low ($g_2(0)=0.22$), which attests to the single-photon character of the emission. The photon coincidence correlation can also be used to determine the lifetime of the emitter state (Figure \ref{correlation}b). The lifetime is equal to the characteristic decay time of the antibunching feature in the $g_2(\tau)$ profile at the low photon rate limit. The lifetime obtained in our experiment ($t_{decay}=4.8\,$ns) is in agreement with the results obtained in other studies \cite{srivastava2015, kumar2015dwuwarstwa, he2015, koperski2015, chakraborty2015}.

We also used the photon coincidence correlation experiment to verify the temporal stability of the emitter. Fluctuations of either the emitter intensity or the intensity of its excitation source would result in the effect of photon bunching. We observe no decrease of $g_2$ at large $\tau$, confirming the good stability of the emission center and the $\upmu$-LED. We also observed good long-term stability of our device. During several days of measurements, we have observed no significant variation in the emission efficiency of the selected center.

\section{Conclusion}

In conclusion, we have demonstrated that light emission from TMD monolayers can be efficiently generated when they are directly deposited on, and excited by the (In,Ga)N $\upmu$-LEDs. Our hybrid devices, $100\,\mu$m in size, can be seen as an optional route towards the effective, electrically driven generation of light from TMD semiconductors. The devices equipped with different MoS$_2$, WSe$_2$, WS$_2$ and MoSe$_2$ monolayers have been shown to be operational at low temperatures thanks to the unique design of our (In,Ga)N micro-LEDs. A proof of concept for a new, electrically driven source of single photons has been demonstrated with one of our (In,Ga)N $\upmu$-LEDs/WSe$_2$-monolayer devices.

\section*{Acknowledgements}

This work is supported by
TEAM-TECH POIR.04.04.00-00-210C/16-00 and TEAM
POIR.04.04.00-00-1A18/16-01 (ATOMOPTO) projects of the Foundation for
Polish Science co-financed by the European Union under the European
Regional Development Fund. This work was supported under project 
\\2021/41/N/ST3/04240 by National Science Centre, Poland. A.R. acknowledges support from the "Diamentowy Grant" under decision DI2017/008347 of MEiN of Poland.
The work has been also supported by EU
Graphene Flagship project.

\end{document}